# Compact 200 kHz HHG source driven by a few-cycle OPCPA

Anne Harth,[1,2,+] Chen Guo,[1,+] Yu-Chen Cheng,[1] Arthur Losquin,[1] Miguel Miranda,[1] Sara Mikaelsson,[1] Christoph M. Heyl,[1,3] Oliver Prochnow,[4] Jan Ahrens,[4] Uwe Morgner,[5] A. L'Huillier,[1] and Cord L. Arnold[1,*]

[1]Department of Physics, Lund University, P. O. Box 118, SE-22100 Lund, Sweden

[2]Max-Planck-Institut für Kernphysik, Saupfercheckweg 1, 69117 Heidelberg, Germany

[3]JILA, 440 UCB, University of Colorado, Boulder, USA

[4]VENTEON Laser Technologies GmbH, Hollerithallee 17, D-30419 Hannover, Germany

[5]Institut of Quantum Optics, Leibniz Universität Hannover, Welfengarten 1, D-30167 Hannover, Germany

[+]Both authors contributed equally to this work

[*]Cord.Arnold@fysik.lth.se

Abstract

We present efficient high-order harmonic generation (HHG) based on a high-repetition rate, few-cycle, near infrared (NIR), carrier-envelope phase stable, optical parametric chirped pulse amplifier (OPCPA), emitting 6fs pulses with 9μJ pulse energy at 200kHz repetition rate. In krypton, we reach conversion efficiencies from the NIR to the extreme ultraviolet (XUV) radiation pulse energy on the order of ~$10^{-6}$ with less than 3μJ driving pulse energy. This is achieved by optimizing the OPCPA for a spatially and temporally clean pulse and by a specially designed high-pressure gas target. In the future, the high efficiency of the HHG source will be beneficial for high-repetition rate two-colour (NIR-XUV) pump-probe experiments, where the available pulse energy from the laser has to be distributed economically between pump and probe pulses.

1. Introduction

Attosecond pulses in the extreme-ultraviolet (XUV) spectral range are generated via high-order harmonic generation (HHG), when intense ultrashort laser pulses are focused into a generation gas. The interaction of attosecond pulses with atomic, molecular, or solid state systems in pump-probe experiments may result in the creation, measurement, and ultimately control of ultrafast electronic wave packets in matter [1]. Such advanced experiments are enabled by state-of-the-art laser technology. Since the first demonstration of attosecond pulses in 2001 [2,3], attosecond pump-probe experiments have mostly been performed employing titanium-doped sapphire (Ti:Sa) chirped pulse amplification (CPA) lasers with repetition rates in the range of a few Hz to a few kHz and pulse energies in the mJ range. While some experiments work in a multi-event regime, where the signal-to-noise ratio can be increased by detecting many interaction events per laser shot, different types of experiments require few events, ultimately just a single event, per shot and thus strongly benefit from high-repetition rates. This is the case, when for example correlated electrons and ions are detected in coincidence with the goal of unravelling particular ionization pathways or when the detection scheme is affected by space-charge effects, like in time-resolved photo-electron microscopy (PEEM) or surface-based photo-electron spectroscopy [4]. In both cases, event rates of one event per laser shot or less are required. In combination with low repetition rate laser sources the acquisition time to build up significant statistics is then inconveniently long and may often exceed the time frame for which such sources can be kept sufficiently stable. For these experiments, attosecond sources with a repetition rate higher than 100 kHz are very advantageous.

Different laser technologies for driving HHG at high-repetition rate systems were demonstrated so far, e.g. based on intra cavity HHG [5], Yb-doped fibre amplifiers [6], thin-disk oscillators [7], few-cycle Ti:Sa lasers [8], or few-cycle optical parametric chirped pulse amplifier (OPCPA) systems [9,10]. While there is no fundamental reason for the HHG conversion efficiency to drop with decreasing laser pulse energy and tighter focusing [11], efficient HHG with a pulse repetition rate of >100kHz and driving pulse energy in the µJ range still is experimentally challenging. This is mostly attributed to the complexity of the generation gas delivery system, which must provide localized high pressure (up to some bars) in the interaction region, while not contaminating the vacuum in the chamber. Resulting from these challenges, to the best of our knowledge, no attosecond pump-probe experiments have so far been performed at repetition rates exceeding a few ten kHz.

In this work, we present a high-repetition rate, HHG based, XUV, attosecond pulse source, driven by a few-cycle, carrier-envelope phase (CEP) stable, OPCPA laser. Efficient HHG in neon, argon and krypton is demonstrated and the source is very promising for realizing high-repetition rate XUV-IR attosecond pump-probe experiments. In krypton, HHG can be driven with pulse energy as low as 2.5µJ, which allows splitting off a significant fraction of the driving pulse energy for the probe arm of a pump-probe interferometer. This was made possible by a rigorous optimization of the driving pulses and a newly designed HHG gas target for improving phase-matching. We present a detailed three-dimensional characterization of the OPCPA output pulses using the dispersion scan (d-scan) technique [12] and spatially-resolved Fourier transform spectrometry [13,14], which enables us to minimize spatiotemporal couplings, often originating from (misaligned) non-collinear parametric amplifiers. We describe our HHG setup, including the high-pressure nozzle design, measurements of the photon-flux, beam profile, and XUV spectra for HHG in neon, argon and krypton.

2. High-repetition rate optical parametric chirped pulse amplifier

The two-stage non-collinear optical parametric chirped pulse amplifier is sketched in Fig. 1. It can be divided into three sub-units: a CEP stabilized Ti:Sa oscillator, an Yb-doped fibre CPA laser, and a two-stage, non-collinear parametric amplifier (NOPA).

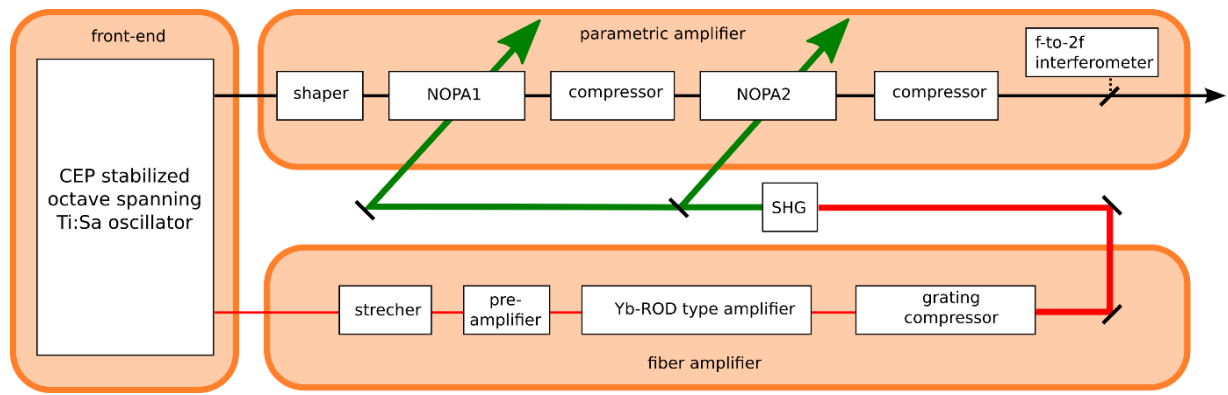

Figure 1: Schematic setup of the non-collinear, two-stage OPCPA: The front-end is an octave spanning Ti:Sa oscillator which seeds a fibre amplifier and a parametric amplifier. The frequency doubled fibre amplifier output provides the pump for the parametric amplifier stages.

2.1 Oscillator

The CEP stabilized octave-spanning Ti:Sa oscillator (VENTEON, 80MHz, 6fs) is the front end of the OPCPA. Its spectrum is used to seed both the parametric amplifier and the fibre CPA, inherently optically synchronizing the two. The frequency doubled output of the fibre CPA serves as pump for the parametric amplifiers.

2.2 Fibre amplifier

A narrow fraction of the oscillator spectrum around 1030nm is selected with a dichroic mirror to seed the fibre amplifier chain, which consists of a chirped fibre Bragg grating for stretching, a fibre preamplifier, two parallel rod-type amplifiers, and two grating compressors. In the preamplifier, the seed is amplified to 600mW and the repetition rate is reduced from 80MHz to 200kHz with the help of three amplification stages and two pulse pickers. The output of the preamplifier is divided into two pulses with the same power to seed two Yb-doped rod-type amplifiers in parallel (NKT-DC-285/100-PM-Yb-ROD). The rod-type amplifiers deliver 150µJ pulse energy before compression. With an 80% efficiency of the grating compressor, the output of the fibre amplifiers is 24W each. Thermal management and appropriate cooling is essential to obtain high power and good mode quality from the rod-type amplifiers. Note that for the results presented in this work only one rod-type amplifier was used.

The pulse duration of the fibre CPA output pulses is approximately 350fs, measured with an auto-correlator assuming a Gaussian-correction factor. For second harmonic generation the output of the fibre amplifier is focused into a 1mm long type-I β-barium borate (BBO) crystal, resulting in 11W at 515nm, corresponding to 48% conversion efficiency. The pulse duration of the second harmonic is estimated to be approximately 250fs. By using a λ/2-plate and a thin-film polarizer (TFP) the second harmonic is divided into two beams to pump two NOPAs.

2.3 Parametric amplifier

The parametric amplification part is comprised of a pulse shaper, two NOPAs built in the Poynting-vector walk-off compensation geometry, chirped mirrors for intermediate and final compression, and a f-to-2f interferometer for slow drift control of the carrier-envelope phase (CEP).

2.3.1 Pulse shaper

In order to reach a pulse duration close to the Fourier limit after the parametric amplification, the main part of the oscillator output (640nm- 1100nm) is sent through a homemade spectral phase-shaper, which is designed in the typical 4f-geometry (e.g. [15]) using reflection gratings (200 lines/mm) and spherical mirrors (f = 660mm). In the Fourier plane a single display spatial light modulator (SLM, Jenoptik) with 640 pixels controls the spectral phase of the incoming light. While the shaper itself should be dispersion free, the positive dispersion corresponding to approximately 7 m of air and the SLM material will stretch the sub-7fs pulse. This is compensated by three reflections on a double-chirped mirror pair and a Calcium Fluoride (CaF$_2$) wedge pair. In order to match the beam size in the Fourier plane to the pixel size of the SLM (100 µm), a mirror telescope enlarges the incoming beam radius to 2 mm. After the recombining grating, a second telescope decreases the beam size back to the original size and the beam is sent to seed the first NOPA. The transmission through the shaper setup is 27%.

2.3.2 NOPA1

The pump for the first NOPA stage, 3W, is focussed into a 2 mm type-I BBO (θ = 26.6˚) crystal (beam diameter is approximately 350 µm), resulting in 125 GW/cm$^2$ peak intensity. The seed pulse diameter in the BBO crystal is slightly smaller. For this NOPA stage the non-collinear angle between pump and seed is approximately α= 2.4˚, ensuring broadband amplification. A motorized translation stage in the seed path compensates for temporal drifts between the pump and the seed (more information about the delay stabilization is given in section 3.4).

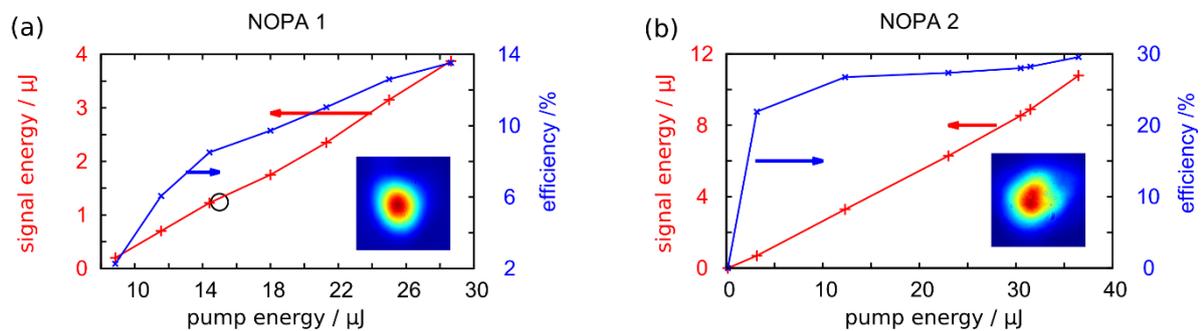

Figure 2: Energy slope of NOPA1 (a) and NOPA2 (b) and respective far-field beam profiles. The red data points were obtained by measuring the average power of the full beam. The blue curve represents the efficiency of the amplification process. The small circle in Fig. (a) marks the working point of the first NOPA.

The first NOPA amplifies the 10mW (80MHz) seed to 260mW (200kHz), which corresponds to an efficiency slightly below 10%. In Fig. 2(a) the output energy of NOPA1 is plotted versus the pump energy. The red data points were obtained by measuring the average power of the full beam. Above a pump energy of 15µJ, the amplified signal spectrum starts to show undesired parasitic effects. At that point, the efficiency curve changes its slope. The black circle marks the chosen working point for NOPA1. With only 3W pump power, NOPA1 works in a very relaxed condition, which leads to good temporal and spatial properties. The inset in Fig. 2(a) shows the measured beam profile. Spatial and temporal characterization of the first NOPA stage are shown in section 3. The dispersion of a 2 mm thick BBO crystal stretches a compressed <7fs pulse to approximately the length of the pump pulse. Thus, the seed pulse must be recompressed before NOPA2. Two reflections from a double-chirped mirror pair (Venteon DCM7) slightly over-compensate the dispersion; this is balanced by 2 mm of fused silica and about 60 cm of air between the NOPA stages.

2.3.3 NOPA2

For the second NOPA stage, the signal is focused to a diameter of 380µm into a similar type-I crystal as for the first NOPA (2mm thick). With a pump pulse diameter of approximately 450µm and a power of 7.3W, the intensity in the crystal is 180GW/cm$^2$. Also the non-collinear angle between pump and signal is similar to the first NOPA setup. The pump-energy to signal-energy output slope for NOPA2 is shown in Fig. 2(b). The efficiency curve flattens very early at a signal energy of a few µJ, indicating that at higher pump energies the NOPA stage is working in saturation conditions. At the same time parasitic effects start altering the spectrum. The decision to nevertheless pump the second NOPA with 7.3W (36.5µJ) is based on a spatio-temporal characterization of the output (presented in section 3.3), showing that the distortions to the amplified spectrum and the spectral phase are still acceptable. With a pump power of 7.3W the signal power reaches 2.3W, which corresponds to 11.5µJ. If the seed from the oscillator is fully blocked the measured incoherent super-fluorescence after the second stage is 237mW which is 10% of the output power. After the second NOPA, there are two chirped mirror compressors. The first one, with four reflections on a double-chirped mirror pair (Venteon DCM7) compensates the dispersion from NOPA2. A beam splitter placed directly behind the compressor sends 5% of the power to an f-2f interferometer for CEP-stabilization (see section 3.4.2). The second chirped mirror compressor, with seven reflections on the mirror pair (Venteon DCM11) pre-compensates the significant amount of dispersion in our beamline on the way to the HHG setup, originating from i.e. the glass wedges for the d-scan, vacuum entrance windows and the focusing lens for HHG. The OPCPA output pulse energy is measured directly behind the pre-compressor resulting in 9µJ.

3. Characterization

3.1 Spectrum

The spectral power at different stages in the system is shown in Fig. 3. The spectrum originating from the oscillator (grey line), seeding the first NOPA stage, is slightly modified by the reflectivity of the gratings in the shaper and furthermore cut at the edges due to the hard aperture of the LCD display. The transmission through the hard aperture is chosen in order to match the amplification bandwidth of the parametric amplifier, i.e. 640nm to 1100nm (or 270 THz to 470 THz).

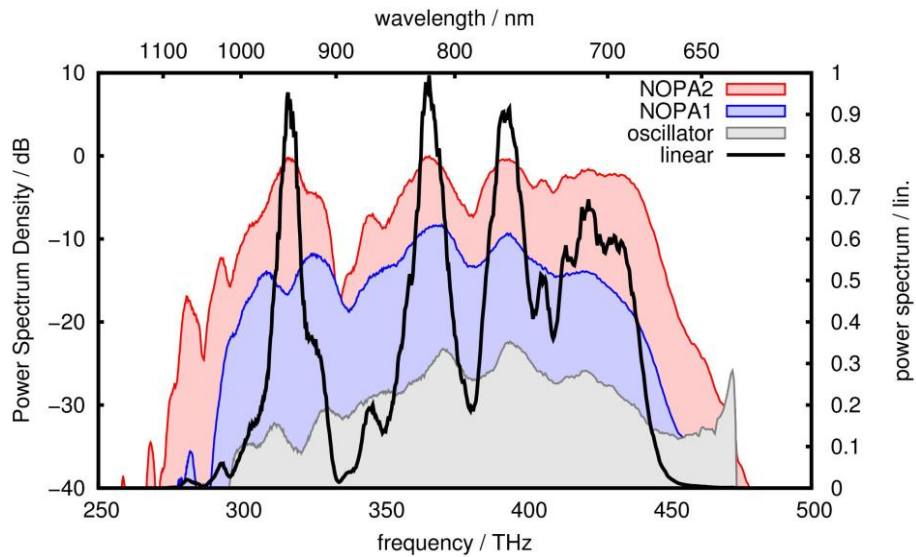

Figure 3: Seed spectrum to the first NOPA stage (grey), spectrum after the first NOPA (blue), and the output spectrum in logarithmic (red) and linear scale (black).

The first NOPA covers almost the whole bandwidth of the seed and amplifies it from 0.125nJ to 1.5µJ per pulse. Apart from the well-known parasitically phase-matched SHG around 340THz [16] in the Poynting-vector walk-off compensation geometry, the shape of the spectral amplitude is maintained in the parametric amplification process, reflecting the absence of strong parasitic and undesired wave mixing processes. The second stage also covers the whole bandwidth of the seed spectrum (see red curve in Fig. 3), but shows first signs of nonlinear distortions, e.g. strong amplification at the edges of the spectrum, a strongly modulated structure around 440THz as well as new peaks and dips around 300THz and the filled gap at 320THz [17]. The black curve presents the output spectrum in a linear scale.

3.2 Temporal characterization

The duration of the NOPA output pulse is characterized with the d-scan technique [12]. Since this method requires the pulses to be focussed into a nonlinear second harmonic generation crystal, we simulate the focus conditions, which are present in the HHG setup. For reasons discussed in section 4, we focus the pulses with an achromatic lens. The NOPA pulses are therefore characterized not behind the first compressor after the two NOPA stages (see section 2.3.3), but behind the second chirped mirror compressor, a BK7-wedge pair as well as a vacuum window and an (identical to the HHG setup) acromatic focussing lens. Fig. 4 (a) shows the measured and reconstructed d-scan traces without an additional phase applied to the phase shaper. The fundamental spectrum was measured with a calibrated optical spectrum analyser (OSA, Andor). The tilted shape of the d-scan traces indicates that the different spectral components of the broadband pulse cannot be compressed simultaneously.

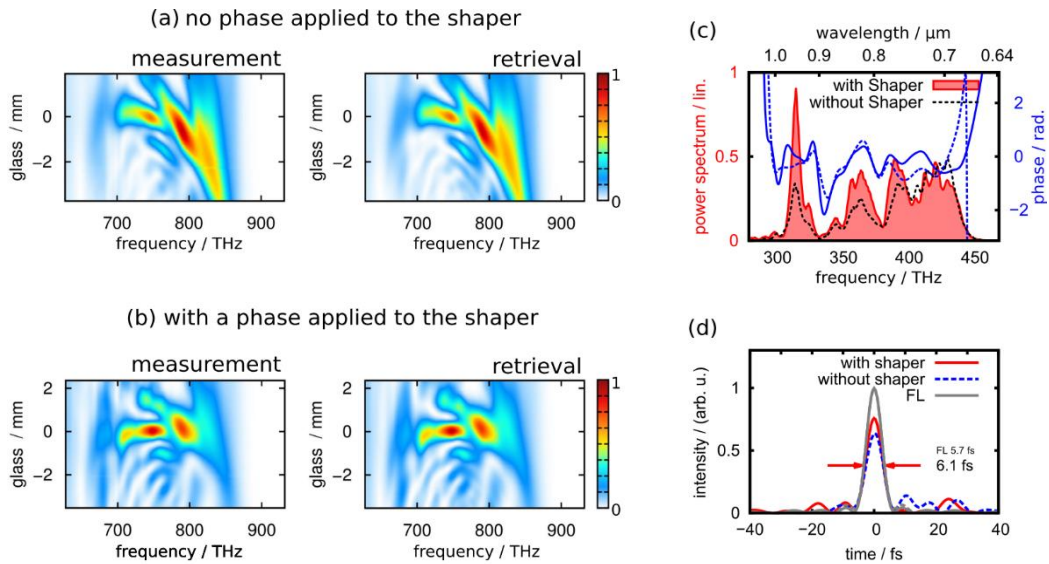

Figure 4: Measurement of the laser output pulse duration via d-scan. (a) measured and retrieved d-scan traces without a phase applied to the shaper. (b) d-scan traces with a phase applied to the shaper. (c) measured spectrum and spectral phase with (red filled spectrum, blue solid curve) and without (black and blue doted curve) a phase applied to the shaper. (d) Reconstructed pulse durations without applied phase (blue dotted line), with applied phase (red solid line), and for comparison the Fourier-transform-limit (grey line).

Fig. 4(c) shows the spectral power (dotted black line) and phase (dotted blue line). The pulse suffers from ringing in the spectral phase, originating from uncompensated dispersion components, e.g. from double chirped mirrors. The reconstructed temporal profile is shown in Fig. 4(d) (blue dotted curve, 6.3fs pulse duration). It is obvious that the pulse cannot properly be compressed, which is mainly reflected by the fact that the peak power is much lower than for the Fourier-transform-limited case (grey line).

Because the pulse shaper is placed before the first parametric amplifier, the dynamic range of the phase that can be applied is limited. Applying the exact spectral phase, necessary to obtain a Fourier-transform-limited pulse behind the two NOPA stages at the HHG focus, is unfortunately not possible, because it would significantly change the temporal shape of the seed pulses for NOPA's. Consequently, the pulse durations of the pump and seed pulses would not be matched anymore, which will reduce the efficiency of the parametric amplification. As a compromise, a spectral phase is applied, which cleans the output pulses from the OPCPA, i.e. leading to less satellite pulses and more power in the peak of the pulse, while not impacting the parametric amplification too much. The measured and reconstructed d-scan is shown in Fig. 4(b). The d-scan trace is not tilted anymore, indicating that the pulse can be better compressed temporally. The spectrum (Fig. 4 (c), filled red curve) supports a Fourier-transform-limited pulse duration of 5.7fs. With the shaper, the pulse could be compressed to 6.1fs, (Fig. 4 (d), red line) and the pulse peak power could be increased to 75% of the Fourier-transform-limited power. Taking this into account, we conclude that of the 9µJ pulse energy emitted by the laser, about 6.75µJ are contained in the main pulse, resulting in a peak power of 1.1GW.

3.3 Spatio-temporal couplings

In this section, we present a spatio-temporal characterization of the output pulses from the OPCPA. Optical parametric amplification naturally introduces spatio-temporal coupling, which can be kept small in the so-called *magic angle configuration* [16,18]. A misaligned NOPA stage increases distortions to the spatial and the spectral phase and leads to stronger spatio-temporal couplings, which impact the performance of subsequent NOPAs and ultimately reduce the intensity to which the OPCPA output can be focused. Therefore, spatio-temporal characterization is a very powerful tool to measure and understand the impact of phase distortions and spatio-temporal couplings and to avoid them as much as possible during the amplification process in a NOPA.

As mentioned in section 3.1, the amplified spectrum after the second NOPA stage (see Fig. 3) shows spectral features, which are not related to the oscillator seed spectrum. This indicates the presence of parasitic effects during the amplification process. One prominent parasitic effect is the second harmonic generation of a fraction of the signal (and the idler) in the Poynting-vector walk-off compensation geometry [16], which leads to distortions of the spectral amplitude and phase of the signal (and idler). Furthermore, the second NOPA stage operates in saturation for some spectral parts which might lead to back-conversion to the pump frequency and local distortions to the spectral phase and amplitude. This will affect the compressibility of the output pulse [17,18]. Since in a non-collinear optical parametric amplifier spatial distortions are expected as well, originating e.g. from (i) the angular-frequency dependent nonlinear gain distribution, (ii) a pulse front miss-match of the pump and the signal beams [18,19] and (iii) from the birefringent nature of the amplification crystal (BBO), temporal distortions are strongly coupled to the spatial coordinates of the beam.

In contrast to previous spatial and temporal characterization techniques applied to high-repetition rate NOPAs based on e.g. SPIDER measurements taken at different positions of the beam [20], imaging spectrometry [20] or spatially resolved spectral interferometry (SRSI) [21], we apply a homemade technique based on spatially-resolved Fourier transform spectrometry, capable to measure the spatio-spectral phase over large apertures [13,14]. With this technique, we can access the full spatio-temporal information of the pulse regarding the amplitude and phase in space and time at arbitrary planes. A similar approach based on the SEA-F-SPIDER technique was recently applied to the characterization of another high-repetition rate, few-cycle OPCPA laser [22,23].

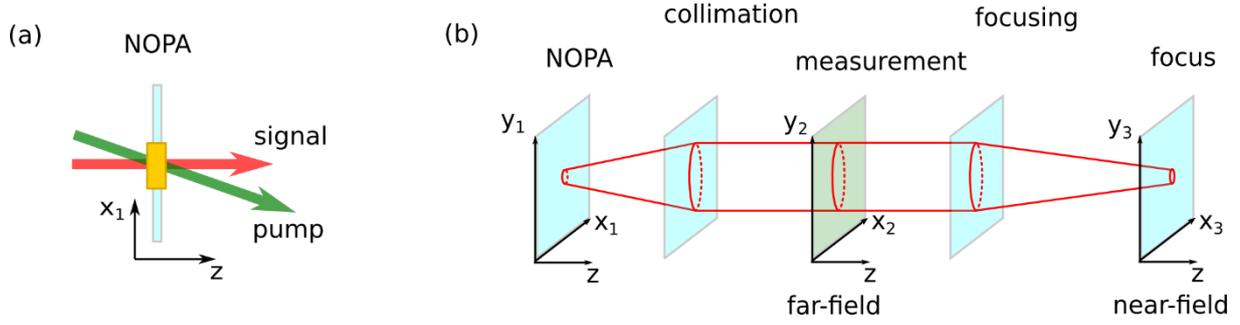

Figure 5: (a) orientation of the pump (green) and signal (red) beams in a NOPA geometry, (b) schematic view of the different relevant planes mentioned in the text: The NOPA, the source of spatio-temporal coupling; the far-field, the plane of measurement after collimation; the near-field, the focus in the HHG setup or a subsequent NOPA stage.

Spatio-temporal couplings originating from NOPAs employing uniaxial crystals, e.g. BBO, are expected to occur in the plane, where the seed and pump beams are non-collinearly mixed, i.e. the walk-off plane or x-z plane, if the z-direction is the propagation direction of the pump (see Fig. 5(a)). In the y-z plane, no strong spatio-temporal distortions are expected and observed in our analysis. The most prominent spatio-temporal distortion, originating from NOPA stages detuned from the magic angle configuration, should be angular dispersion, which also causes pulse front tilt and potentially spatial dispersion in the far-field. For a detailed discussion of spatio-temporal couplings and their different representations we refer to an article by Akturk and co-workers [24]. The analysis of our measurements concentrates on distinguishing the following spatio-temporal coupling contributions.

- spatial dispersion where the coupling $\gamma$ is found in the amplitude

$$E(x + \gamma_{x,\omega}, y + \gamma_{y,\omega}, \omega) e^{i\,\varphi(x,y,\omega)}$$

- angular dispersion where the coupling $\gamma$ is found in the phase

$$E(x,y,\omega) e^{i\,\varphi(x+\gamma_{x,\omega}, y+\gamma_{y,\omega}, \omega)} \to \tilde{E}\left(k_x + \tilde{\gamma}_{k_x,\omega}, k_y + \tilde{\gamma}_{k_y,\omega}, \omega\right) e^{i\,\tilde{\varphi}(k_x,k_y,\omega)}$$

Here, $k_x$, and $k_y$ are the transverse wavenumbers in x- and y-direction, respectively, which can be obtained by Fourier transforming the electric field $E$ of the pulse along the space coordinates. Angular dispersion, such that different frequencies propagate into different directions, will be visible as asymmetry (tilt) in a ($k_{x,y}$-ω)-representation of a pulse. Spatial dispersion, such that the frequency content of the pulse is not evenly distributed in space, will be visible as asymmetry in a (x,y-ω)-representation.

For our spatio-temporal analysis, we distinguish three different planes, which are illustrated in Fig. 5(b), i.e. the NOPA plane ($x_1$, $y_1$), where the couplings originate from, the far-field ($x_2$, $y_2$), where we perform the characterisation and the near-field ($x_3$, $y_3$) which is obtained by refocusing the pulses either with the purpose of generating high-order harmonics or to couple to an additional NOPA. Our characterization technique measures the phase and amplitude of the pulse $E(x_2, y_2, \omega)$ in the far-field [13], i.e. after collimation at some distance away from the BBO crystals. By simple Fourier operations, we can obtain the pulse in any other representation, i.e. ($x_2$,$y_2$-t) and ($k_{x2}$,$k_{y2}$-ω), and by numerical propagation also in any other plane, e.g. in the near-field at the focus of our HHG setup.

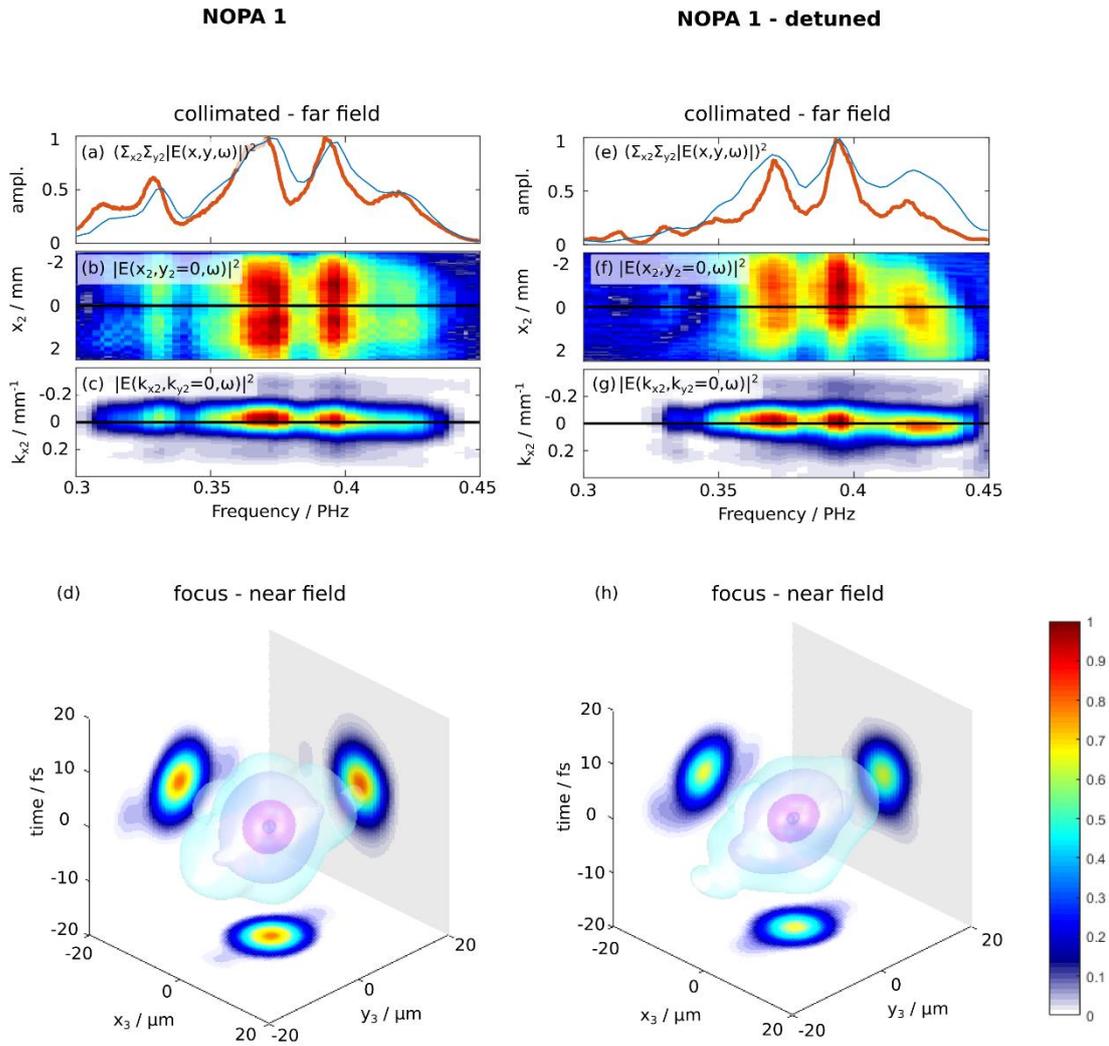

Figure 6: Spatio-temporal characterization of the pulse, after NOPA 1. Shown is only the x-direction since no strong distortions are expected in the y- direction. (a)-(d) best effort aligned NOPA stage, (e)-(h) detuned NOPA stage. (a) and (e) averaged spectrum (blue) compared to a spectrum measured by a calibrated spectrum analyser; (b) and (f) cuts of the intensity in ($x_2$-ω)-representation at $y_2$=0; (c) and (g) cuts of the intensity in ($k_{x2}$-ω)-representation at $k_{y2}$=0; (d) and (h) calculated intensity of the pulses in the near-field in ($x_3$,$y_3$-t)-representation.

We start by characterizing the spatio-temporal couplings originating from the first NOPA stage (NOPA1), which is built with very relaxed pumping conditions. The measurement is performed after collimating the diverging beam from NOPA1, i.e. in the far-field. We characterize two different conditions of NOPA1: first, it is aligned to the best of our abilities and second, we deliberately detuned the angle between the pump beam and the crystal axis in order to generate angular dispersion and pulse front tilt as well as spatial dispersion in the far-field. For the first condition, we optimized the NOPA stage by carefully choosing the pump conditions regarding the pump beam size and the power, by that making sure that this stage works far from saturation (see section 2.3.2). This can be tested by a pump-signal efficiency curve and by avoiding distortions in the amplified spectrum. At the same time, the non-collinear angle is optimized for broadband amplification. Fig. 6 summarizes the extracted data from the spatio-temporal characterization of NOPA1. Fig. 6 (a) shows the spatially integrated spectral power (blue line), (b) the intensity in ($x_2$-$\omega$)-representation (i.e. a cut for $y_2$=0), and (c) the intensity in ($k_{x2}$-$\omega$)-representation (cut at $k_{y2}$=0) for the well aligned NOPA1 on the left side and with deliberate misalignment on the right side ((e)-(g)). Also shown in Fig. 6 (a) und (e) is the spectrum measured with the optical spectrum analyzer (red line). The spectra agree well in both cases, which is a very important confirmation that our Fourier transform spectrometry approach gives reliable data. For the well aligned NOPA no distinct asymmetry is observed neither in the ($x_2$-$\omega$)- nor the ($k_{x2}$-$\omega$)- representation of the measured pulse, indicating that there is no significant spatio-temporal coupling. The situation is different when NOPA1 was deliberately misaligned. Fig. 6(f) shows a clear asymmetry of the pulse in the ($x_2$-$\omega$)-representation, i.e. spatial dispersion originating from angular dispersion in the NOPA plane. This is accompanied by a slight tilt of the angular spectrum, which however is less obvious Fig. 6(g).

We investigate the focusability by numerically propagating the pulses for the well-aligned and misaligned NOPA1 to the near-field, assuming aberration-free and dispersion-free focussing conditions. We follow a procedure, outlined by Giree and co-workers [18], where the spectral phase is not explicitly considered and where the pulses are assumed temporally transform-limited on-axis. Off-axis, however, the spectral phase can be more complicated, as the phase difference between the on-axis and off-axis phases is kept in this analysis. This allows us to disentangle the impact of the introduced spatio-temporal couplings on the focusability from the impact on the pulse duration. For the numerical propagation from the far-field to the near-field an angular plane wave decomposition is used [25], providing the pulses in ($x_3$,$y_3$-$\omega$)-representation (see Fig. 5(b)). A Fourier transform in the frequency direction, finally gives the focused pulses in ($x_3$,$y_3$-t)-representation, from which we can extract the maximum intensities. The focused pulses are illustrated in space and time for the well-aligned and misaligned NOPA in Fig. 6 (d) and (h), respectively.

While the differences are subtle in visual comparison, the impact of the misalignment becomes obvious when comparing the maximum intensities with each other as well as with an idealized case for which we numerically removed all spatio-temporal couplings (see [18] for a discussion of the procedure). This can be understood as a three-dimensional Strehl ratio, relating the actual focal intensity, compromised by phase distortions and spatio-temporal couplings, to the focal intensity of a pulse with identical pulse energy, identical spatial profile, but clean spatial phase. For the sake of easy comparison, we normalize the pulses from the well-aligned and deliberately misaligned NOPA1 to the same pulse energy, ignoring the fact that the misaligned NOPA1 should give less pulse energy. By this procedure, we obtain a Strehl ratio of approximately 0.81 for the well-aligned NOPA1, while the value for the misaligned NOPA1 is approximately 0.63. Taking into account the strongly nonlinear intensity dependence of high-order harmonic generation, such difference in focal intensity can largely impact the HHG signal and its conversion efficiency. Furthermore, spatio-temporal couplings might impact the performance of subsequent NOPA stages.

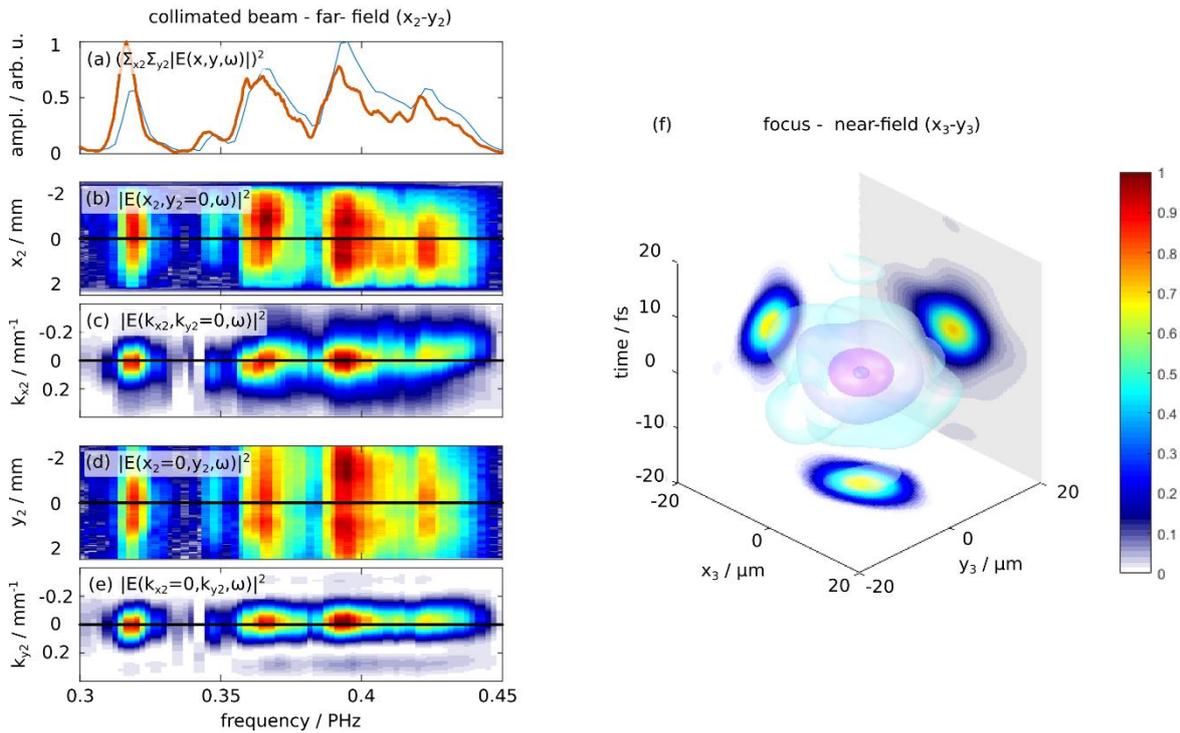

Figure 7: Spatio-temporal characterization of the pulse, after NOPA2: (a) averaged spectrum (blue) compared to a spectrum measured by a calibrated spectrum analyzer; (b) and (d) cuts of the intensity in ($x_2$-$\omega$)-representation ($y_2$=0) and ($y_2$-$\omega$)-representation ($x_2$=0); (c) and (e) cuts of the intensity in ($k_{x2}$-$\omega$)-representation ($k_{y2}$=0) and ($k_{y2}$-$\omega$)-representation ($k_{x2}$=0). (f) Intensity of the calculated pulse in the focus in ($x_3$,$y_3$-t)-representation.

Finally, we also characterized the pulses after the second NOPA stage. Again, the characterization was performed in the far-field after collimation. For this measurement, both the first and the second NOPA were aligned to the best of our abilities. Fig. 7 summarizes the results. The averaged spectra measured with the OSA (red line) and the Fourier transform approach (blue line) are in good agreement again (Fig. 7(a)). In the two middle ((b) and (c)) and the two bottom rows ((d) and (e)) we plot the intensity in $(x_2,y_2-\omega)$- and $(k_{x2},k_{y2}-\omega)$-representations, respectively. Spatio-temporal distortions should be expected in the direction of non-collinear wave mixing in the NOPAs, i.e. the x-direction. This is confirmed by the characterization. In Fig. 7(b) a slight asymmetry in the spectral distribution along the $x_2$-direction is observed, i.e. a sign of spatial dispersion, as well as a slight tilt in the $(k_{x2}-\omega)$-representation (Fig. 7(c)), i.e. a sign of angular dispersion. In the y-direction (Fig. 7 (d) and (e)), no distinct asymmetries can be observed, neither in the $(y_2-\omega)$- or $(k_{y2}-\omega)$-representations of the intensity, indicating that the pulse should focus cleanly in the y-direction. The difference in the focusability of the x- and y-directions becomes obvious, if we interpret the $(k_{x2}-\omega)$- (Fig. 7(c)) and $(k_{y2}-\omega)$-representations (Fig 7(e)) as rescaled versions of the near-field $(x_3-\omega)$- and $(y_3-\omega)$-representations. This is valid if we assume that the far-field and the near-field are connected via their spatial Fourier transforms, i.e. that diffraction in the Fraunhofer regime is valid [25]. From this comparison, it is clear that the transverse size of the focus in the near-field, e.g. the HHG interaction region, is larger in the x- than in the y-direction. This is confirmed by a three-dimensional illustration of the spatio-temporal intensity distribution in the focus, obtained from numerical propagation, which is shown in Fig. 7(f). The obtained Strehl ratio is approximately 0.71.

If we now additionally consider the measured spectral phase of the pulses reflecting the temporal pulse shape at the HHG interaction region, as discussed in section 3.2, only about 75% of the total pulse energy is contained in the main pulse. This additionally reduces the focused intensity compared to an undisturbed pulse, which is transform-limited both in space and time. Behind both NOPAs, including the effect of the spectral phase we obtain a total Strehl ratio of 0.53. The reduction of the peak intensity should have considerable impact on the efficiency of HHG. Please note that even for a carefully aligned, few-cycle OPCPA, as demonstrated in this section, the maximum focused intensity is barely half the possible intensity (not including the spatial-temporal aberration coming from a focussing element). This underlines the importance of spatio-temporal characterization techniques in order to optimize the alignment of optical parametric amplifiers.

3.4 Stability

3.4.1 Active delay stabilization

For long-term stable output power of the OPCPA laser, active delay stabilization between the pump and the seed/signal is essential. As a result of the short pump pulse duration, in our case about 250fs, small fluctuations of the relative timing of the two beams will translate to the amplified output spectrum, the central frequency, the power, and the CEP. The active delay stabilization is based on recording the frequency-doubled, angularly dispersed idler spectrum as emitted from the first NOPA, with a camera. The frequency-doubling originates from a parasitic phase-matching effect. A typical measurement is shown in Fig. 8(a).

If the timing between pump and signal in the crystal changes, the position of the centre of mass of the idler spectrum on the camera chip will move. The active delay stabilization continuously computes the centre of mass, compares it to a saved reference, and generates an error signal, which is used to move a translation stage that controls the delay of the seed pulses to the pump. This stabilization system, running at approximately 5Hz, mostly corrects for thermal path length differences originating from the fibre laser part, which in total is some tens of meters long. Since the largest contribution to the delay instability is addressed with the active stabilization of the first NOPA stage, an active stabilization for the second NOPA is not necessary. With the active delay stabilization, the standard deviation of the average output power over 2 hours is less than 0.5%, see Fig. 8(b). The shot-to-shot energy stability over two hours is 2% rms.

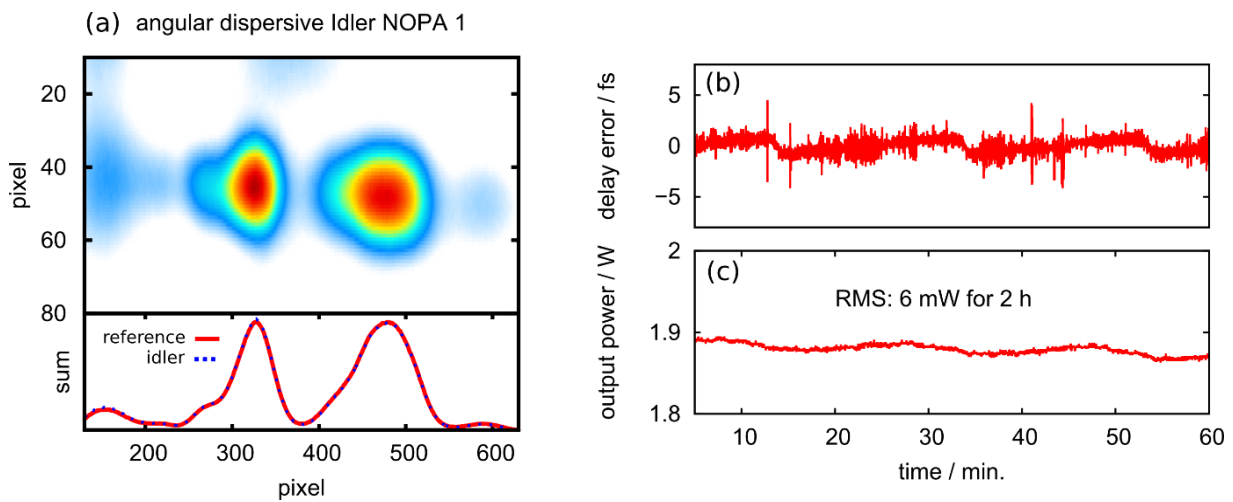

Figure 8: (a) Angularly dispersed idler recorded with a camera close to NOPA 1. The Signal is integrated over the vertical pixel row. The drift of the centre of mass of the summed idler spectrum is used as feedback to control the delay. (b) Delay stabilization error signal to compensate the drift between the pump and the seed pulses. (c) Output power (shown only for 1 hour but measured over 2 hours).

### 3.4.2 CEP-stability

For sub-7 fs pulses around 800 nm the CEP plays an important role. The carrier-envelope frequency of the oscillator is stabilized to a quarter of the repetition rate. The repetition rate of the OPCPA chain is reduced by a factor of 400 to 200 kHz.

Thus, after the first NOPA all amplified pulses should have the same CEP, except for contributions originating from thermal changes, air flow, vibrations, and pump-signal delay drifts. Therefore, a slow CEP feedback loop is implemented. After the chirped mirror compressor behind the second NOPA stage, a broadband beam splitter is used to split off 5% of the total power to measure CEP drift in a f-to-2f interferometer. First, an octave-spanning spectrum is obtained via white-light generation in a 3mm long sapphire plate, second, the red edge of the white-light is frequency doubled (in a 1mm long BBO crystal), third, the spectral interference between the doubled red edge and the blue edge of the white-light is recorded with a fibre spectrometer, and finally the phase of the interference fringes is used to generate an error signal, which is fed back to the oscillator stabilization unit (Menlo Systems). In-loop measurements of the CEP with an integration time of 1.5 ms (or 20 µs) show an rms instability of less than 150 mrad (or 200 mrad).

With a second f-to-2f interferometer, operating with the same principle, out-of-loop CEP measurements were performed. The CEP was measured with an extremely short integration time of 10 µs (two pulses), resulting in an rms instability of 400 mrad, corresponding to 170 as. The sampling rate for the out-of-loop measurements was 270 Hz. A longer integration time of 1.5 ms (300 pulses) lead to less than 250 mrad rms instability.

### 4. High-order Harmonic Generation

The main motivation to develop the OPCPA laser described in this article is the realization of a high-repetition rate HHG source for experimental schemes which actually benefit from high repetition rate, rather than from high attosecond pulse energy. The HHG setup is shown in Fig. 9. A variable aperture is placed in the beam at the entrance to the vacuum chamber in order to adjust the intensity of the NOPA beam and to shape the beam profile. The beam has a diameter of approximately 5mm at this point and is focused with an achromatic lens ( f=50 mm, THORLABS) into the generation gas, leading to a confocal parameter of 200µm. Taking into account the aberrations introduced by using spherical mirrors with short focal length off-axis, and the high price and complexity in alignment of short focal length off-axis parabolic mirrors, an achromatic lens for focusing is a convenient, easy to use, alternative. The positive material dispersion of this lens as well as other material dispersion, e.g. the vacuum entrance window, is pre-compensated by a double chirped mirror pair (see section 2). The pulse duration in the focus generated by an identical achromatic lens was measured to be <7fs (see section 3.2). A homemade gas nozzle (see section 4.1) delivers different noble gases (neon, argon and krypton) at high pressure to the interaction region, where HHG takes place.

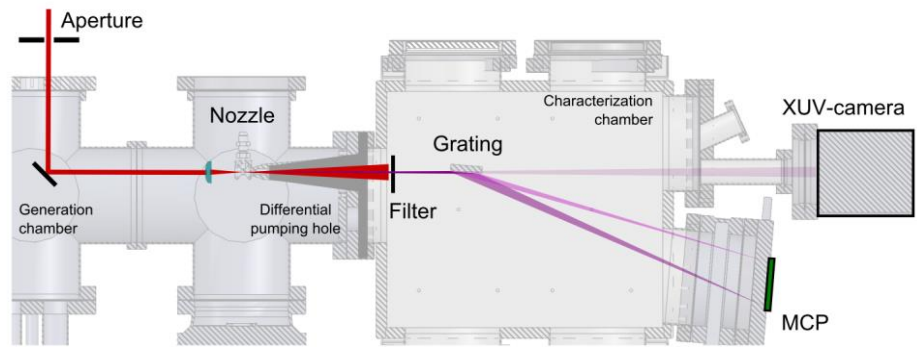

Figure 9: High-order harmonic generation setup, including a variable aperture at the entrance to the vacuum chamber, an achromatic focusing lens, a high backing pressure gas nozzle, a differential pumping hole, metallic filters, an XUV-grating, a MCP, and an XUV-camera.

From the interaction volume, the XUV and IR beams propagate through a differential pumping hole into the characterization chamber. A metallic filter eliminates the fundamental beam and spectrally filters the XUV radiation. It is mounted on a computer controlled rotational holder providing flexibility for changing or removing filters. The homemade XUV-spectrometer is comprised of a grazing incidence, flat-field, XUV-grating, mounted on a rotation and translation stage, and a multi-channel plate (MCP) and phosphor screen arrangement. The phosphor screen is imaged with a camera to record XUV spectra. The rotational stage is used to adjust the angle of the grating. With the translation stage the grating can conveniently be moved out of the beam in order to send it straight on to an XUV-camera for photon-flux and beam profile measurements (see section 4.2).

4.1 Phase matching and nozzle design

The gas supply system is designed to deliver a high gas density to a small medium's length. This is crucial for efficient generation of high-order harmonic radiation in a tight focusing geometry [26].

A sketch of the nozzle system is given in Fig. 10(a). It provides a very localized, high-pressure gas target without contaminating the surrounding vacuum. It is comprised of two nozzles: one for gas injection and one for gas extraction. The injecting nozzle forms a supersonic gas jet through a small, 50μm diameter hole. According to Monte Carlo simulations, computed with DS2V program released with reference [27], the resulting pressure in the interaction volume is estimated to be one order of magnitude lower than the backing pressure. Thus, a backing pressure of 10bar (up to 40bar is possible) approximately results in 1bar (4bar) pressure in the interaction region. The extraction nozzle is connected to a roughing pump, which is supposed to capture most of the injected gas in the generation chamber. The extraction nozzle has a hole of 1mm diameter and is placed 0.2mm away from the injection nozzle. The generation chamber is additionally pumped with a large turbo pump, resulting in a background pressure of below $10^{-2}$mbar, despite the large amount of injected gas.

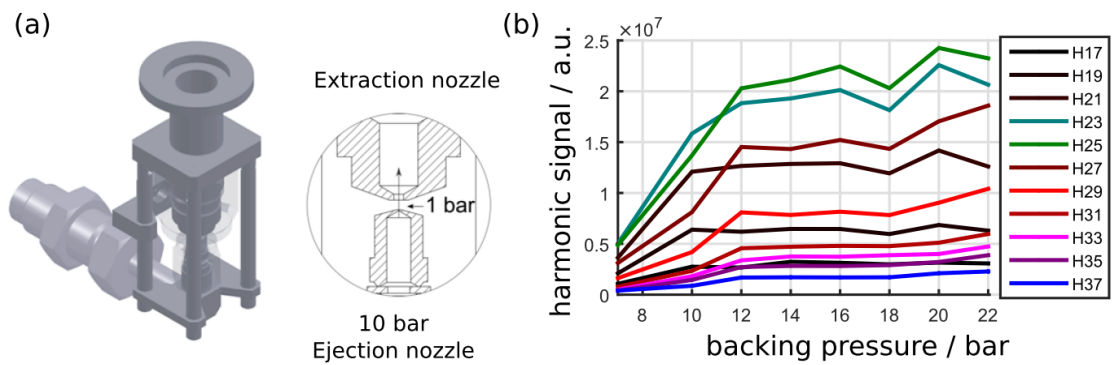

Figure 10: (a) Sketch of the high pressure gas delivery system, including both the nozzle for gas injection and extraction. (b) Pressure scan in Argon.

The simulations suggest that with the current nozzle design the estimated target length approximately 100 µm away from the ejection nozzle, i.e. in the middle between ejection and extraction nozzle, is 200 µm. Outside that length the gas pressure has fallen to below 30%. Thus, the effective gas length is comparable to the confocal parameter of the laser. Fig. 10(b) shows the harmonic yield in argon as a function of the backing pressure. The XUV signal is absorption limited and saturates beyond 12bar. Every time the pressure was changed during the measurement, the nozzle position had to be re-optimized because of a recoil force from the gas jet, which pushes the nozzle away from the laser focus. The dip of the XUV yield at 18bar backing pressure thus can be attributed to a not carefully enough optimized nozzle position.

4.2 Spectra, photon flux and beam profile

4.2.1 XUV spectra

The high-order harmonic radiation is characterized using a homemade XUV-spectrometer consisting of a grating and a MCP (see Fig. 9). The grating (Hitachi, 001-0640) is an aberration-corrected concave grating with variable groove spacing. The XUV beam is simultaneously focused and spectrally dispersed in the horizontal direction, while it is unaffected in the vertical direction. Thus, the vertical direction (not focused) carries information about the divergence of the beam, while the horizontal direction (focused) carries the spectral information.

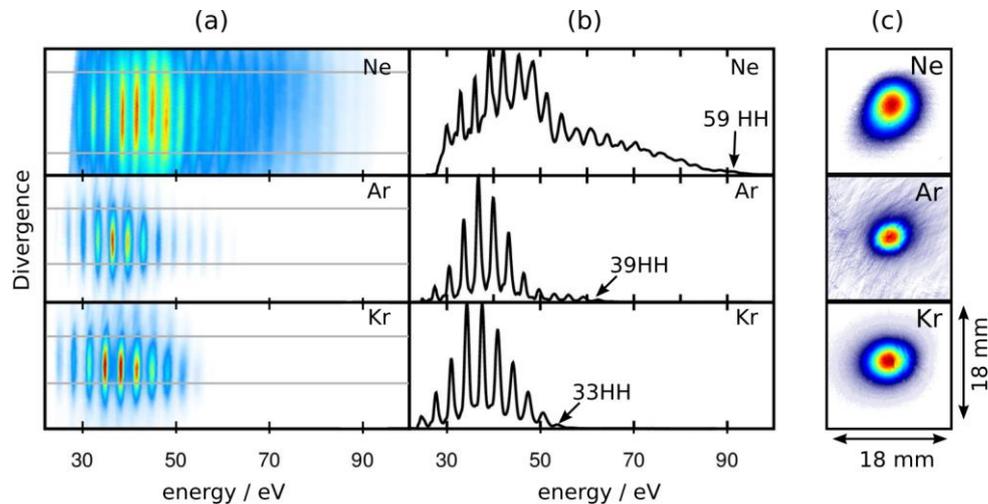

Figure 11: HHG spectra generated in neon, argon and krypton. (a) Divergence of the spectral components of the XUV radiation, (b) Integrated spectra (see grey lines in (a) for the integration limits) (c) XUV beam profiles measured with Al filters and an XUV-Camera.

Fig. 11(a) shows harmonic spectra, i.e. photon energy versus divergence, for neon, argon, and krypton. For all three gases, the nozzle position was optimized for the highest cut-off energy in the high-order harmonic spectra. For argon and krypton contributions from the long trajectories to the HHG spectrum can be identified. Since long and short trajectories differ in their divergence, i.e. the long trajectories are more divergent, their contributions become obvious for large angles. We select the short trajectory contribution to our HHG spectra by integrating in the vicinity of the optical axis. Fig. 11(b) shows the HHG spectra integrated between the two grey lines marked in (a). Except for neon, all other measurements were performed with an aluminium (Al) filter placed before the grating (see Fig. 9). As a result, the low energy sides of the argon and krypton spectra are shaped by the Al-transmission window and by the transmission through a few nm thick oxide-layer, usually covering the filter on both sides [28]. The neon spectrum is influenced around 25eV by the frequency dependent diffraction efficiency of the grating. Krypton shows distinct harmonics from the filter edge around 22eV (H15) to the cut-off at 53eV (H33). In argon, the spectral amplitude decreases around 50 eV due to the presence of the Cooper minimum. The argon HHG cut-off reaches H39 (60eV). For the measurement of the neon HHG spectrum, no filter was used to suppress the strong IR field because the higher absorption edge of Aluminium around 72eV would remove the cut-off of the neon HHG spectrum. The spectrum from 20eV to 40eV overlaps with the second-order diffraction from higher photon energy harmonics. The spectrum from 50eV to 90eV features a continuous structure, which can be attributed to the limited resolution of the spectrometer in this range. Only the presented spectrum in neon was measured with a locked CEP of the amplifier. However, locking the CEP and changing it leads to rich interference pattern in the spectrum of all three gases (not shown here).

4.2.2 Beam profile

The XUV beam profile and the photon flux were measured with a calibrated XUV-Camera (Andor, Ikon-L). In order to protect the camera from the strong IR pulse, all the beam profile measurements were performed with metallic filters. For the beam profile measurement in Neon, shown in Fig. 11(c), one 200nm thick aluminium filter was used. For krypton and we applied two aluminium filters, each 200nm thick. All beam profiles have a homogenous intensity distribution and no significant asymmetry is visible. The FWHM beam sizes after propagation of 87cm (from the generation to the camera) are 5.18mm x 5.74mm for neon, 6.16mm x 5.57mm for argon and 5.09mm x 4.52mm for krypton. Since the beam diameter of the XUV radiation generated in all three gases is approximately 5.4mm at the position of the XUV camera, we can estimate the beam divergence to $\theta_{XUV}$ =2.9mrad. With the above given values for the IR beam the ratio of the IR and XUV beam divergences is $\theta_{IR}/\theta_{XUV}$= 17, corresponding very well to the ratio of their wavelengths, i.e. 800 nm (IR) and 41 nm (XUV).

4.2.3 Photon flux

For the XUV-photon flux measurements, the quantum efficiency for the XUV-camera, the number of generated secondary electrons per absorbed XUV-photon as well as the relative spectral intensity of the XUV radiation were taken into account. The XUV-flux was measured behind aluminium filters in order to avoid exposure of the camera to laser radiation. The values given in Table 1 and 2 are for the bandwidth from approximately 23eV to 73eV.

|  | Neon | Argon | Krypton |
|---|---|---|---|
| IR pulse energy (µJ) | 5.0 | 3.1 | 2.2 |
| Backing pressure (bar) | 30 | 10 | 8 |
| Average Power (µW) | 0.001 | 0.031 | 0.055 |
| Photons / second ($10^{10}$) | 0.16 | 0.50 | 0.95 |
| Pulse energy (pJ) | 0.056 | 0.155 | 0.280 |

Table 1: IR pulse energy used, backing pressure of the gas, XUV average power, XUV photon flux after Al filters and XUV pulse energy for neon, argon and krypton.

Because the transmission of the used aluminium filters is very hard to assess in the experiment, it is difficult to estimate the XUV-flux directly behind the generation (before the filters). While the theoretical transmission of an ideal Al foil with a thickness of 200nm can in principle be calculated, a significant uncertainty remains due to oxide layers forming on the front and back surfaces when the filters are exposed to air, e.g. during mounting. The thickness of these layers can vary between 2nm to 8nm [28] with noticeable impact on the transmission. Therefore, for all quantities given in Table 2, an average 5nm thickness of the oxide-layer is considered.

|  | Neon | Argon | Krypton |
| --- | --- | --- | --- |
| Average power (µW) | 0.04 (± 0.015) | 0.83 (± 0.59) | 1.84 (± 1.36) |
| Photons / second ($10^{10}$) | 0.7 (± 0.26) | 15.4 (± 11.2) | 36.1 (± 27.2) |
| Pulse energy (pJ) | 0.20 (± 0.07) | 4.14 (± 2.94) | 9.19 (± 6.78) |
| IR pulse energy (µJ) | 5.0 | 3.1 | 2.2 |
| Conversation efficiency | 4.0 (± 1.5) x$10^{-8}$ | 1.3 (± 1.0)x$10^{-6}$ | 4.2 (± 3.1)x$10^{-6}$ |

Table 2: Estimated XUV average power, photon flux and pulse energy directly behind the generation for neon, argon and krypton, assuming 5 nm (±3 nm) thick oxide-layers on both sides of the aluminium filters.

With the estimated values for the photon flux, it is possible to calculate the conversion efficiency. In argon and krypton, the conversion efficiency reaches 10e-6. This efficiency does not reach the record of $10^{-5}$ from 1kHz Ti:Sapphire systems [Constant, Willner] but is comparable to state of the art high repetition rate table top XUV sources based e.g. on Yb amplifiers [Cabasse, Rodhardt] or Ti:Sapphire systems [Heyl, Wang] even with such low IR pulse energies.

5. Conclusion

In this work, we demonstrated efficient high-order harmonic generation in neon, argon and krypton driven by a high-repetition rate, few-cycle, NIR, OPCPA laser. The OPCPA and the XUV sources are specially developed for attosecond pump-probe experimental schemes that rather benefit from a rapid iteration of the experiment (i.e. high-repetition rate) than from high attosecond pulse energy. A careful characterization of the OPCPA laser was presented, with particular emphasis on assessing spatio-temporal distortions, potentially arising from misalignment of the non-collinear parametric amplification stages or from parametric amplification being driven in unfavourable conditions. Our measurements underline that even our thoroughly aligned few-cycle OPCPA laser achieves not more than 50% of the theoretically possible peak intensity, when focused. This strongly highlights the importance of spatio-temporal characterization techniques for state-of-the-art laser development.

Despite the low pulse energy, i.e. 9µJ at the laser output, whereas approximately 6.75µJ are contained in the 6.1fs central peak of the emitted pulses, efficient HHG was achieved by the combination of a carefully aligned optical parametric amplifier with a tight focussing geometry including a new generation gas nozzle design. Furthermore, the whole pulse energy was not required to efficiently generate high-order harmonics. Thus, it will be possible in the future to save a fraction of the driving laser pulse energy to be used as probe in attosecond NIR-XUV pump-probe experiments, what to the best of our knowledge remains to be demonstrated for repetition rates exceeding a few kHz. The presented XUV source will enable fundamental attosecond time-resolved pump-probe experiments at high-repetition rates, such as coincidence attosecond spectroscopy and time-resolved photo emission electron microscopy (atto-PEEM), and contribute to increasing the signal-to-noise ratio in attosecond experimental schemes.

Since the driving pulse duration is <7 fs, the generation of single isolated attosecond pulses with this system seems within reach. Currently, three to four half-cycles contribute to our HHG spectra. Thus, for generating single attosecond pulses, additional modification of the driving pulse by gating techniques or further compression will be necessary. This will however additionally reduce the IR peak intensity, again demonstrating the importance of developing efficient HHG sources working at low driving pulse energies.

6.Acknowledgements

This work was partly supported by the European Research Council (Grant PALP), the European Union's Horizon 2020 research and innovation programme under the Marie Sklodowska-Curie grant agreement no. 641789 (MEDEA), the Knut and Alice Wallenberg Foundation, the Swedish Research Council, the Swedish Foundation for Strategic Research and Laserlab-Europe EU-H2020 654148.

7.References